\documentclass[11pt]{article}

\usepackage[inline]{enumitem}
\usepackage{amsmath}
\usepackage{amsthm}

\usepackage[breaklinks=true,
            colorlinks = true,
            linkcolor = blue,
            urlcolor  = blue,
            citecolor = blue,
            anchorcolor = blue]{hyperref}
\usepackage{breakurl}
            
\newcommand{\mailto}[1]{\href{mailto:#1}{\nolinkurl{#1}}}

\title{On decentralized oracles for data availability}

\author{
Jason Teutsch\\
\emph{TrueBit Estsblishment}\\
\mailto{jt@truebit.io}}

\date{December 25, 2017\footnote{This version, updated in 2019, includes minor changes, and corrections.}}

\usepackage{color}
\usepackage{soul}
  \definecolor{lightblue}{rgb}{.60,.60,1}
  
\theoremstyle{definition}
\newtheorem{defn}{Definition}
\newtheorem*{mainloop}{Main Loop}
\newtheorem*{consistency}{Consistency Axiom}
\newtheorem*{upload}{Upload Axiom}
\newtheorem*{directory}{Directory Axiom}

\numberwithin{equation}{section}
\DeclareMathOperator{\hash}{hash}
\begin{document}

\maketitle

\begin{abstract}
Nakamoto consensus, the protocol underlying Bitcoin, has the potential to secure a new class of systems which agree on non-mathematical truths.  As an example of this capability, we propose a design for a trustless, data availability oracle.  This exposition reduces the problem of determining whether or not a registered datum is publicly available to the problem of constructing a network in which either almost all nodes can download a given datum, or almost none of them can.  
\end{abstract}

\section{Silo effect}

Traditional blockchains, like Bitcoin \cite{bitcoin} and Ethereum \cite{ethereum}, live a lonely life.  By design, miners, which constitute the ``eyes and ears'' of each network, have no formal means to reach consensus about external events.  While Oraclize~\cite{oraclize} and Reality Keys~\cite{realitykeys} successfully pipe Internet information into Ethereum, their bridges rely on trusted authorities.  These efficient, oracle methods, and even aggregating methods like Chainlink~\cite{chainlink} suffice for many applications, however they also introduce central failure points in Ethereum's otherwise autonomous, reputationless system.

Unstoppable, decentralized applications, such as the original Livepeer protocol~\cite{livepeerwhitepaper} demand not only trustless computation oracles, like TrueBit~\cite{truebit}, but also un-erasable data storage.  Ethereum provides scarce but immutable ``on-chain'' storage, however prohibitive expense limits on-chain storage to only the most laconic of messages.  Ethereum's security depends on every miner storing every byte of blockchain data until the end of time, hence substantive data uploads on this system remain unwieldy.  Moreover, propagating more than a little data in each block could induce rational miners to skip block verification and even break the underlying consensus, a phenomenon known as the Verifier's Dilemma~\cite{LTKS15}.  Indeed, rational miners might not wait for downloads. We therefore consider off-chain alternatives.

We set ourselves the task of constructing a trustless oracle which can confirm off-chain data availability.  This use case fundamentally differs from that of traditional, cloud storage platforms, such as Dropbox~\cite{dropbox}, Storj~\cite{storj}, Sia~\cite{sia}, and Maidsafe~\cite{maidsafe}, which place users in control of the data that they upload.  Dropbox, in particular, allows uploaders to decide whether and when to share their data with others.  In a data availability scenario, on the other hand, disappearance of data can destroy high-stakes financial transactions as well as critical computations.  The network must therefore counteract users' incentives to \emph{pretend} that they have ``published'' data.  For illustration, a Bitcoin transaction should not magically disappear after the system confirms its funds as spent, and neither should the input for a TrueBit computation vanish before the system has had time to process it. 

To distinguish the present approach from existing projects like Filecoin~\cite{filecoin} and Swarm~\cite{swarm}, we explicitly isolate the data availability problem from other commonly associated concepts such as data sharding, privacy, computation, scalability, and Ethereum.  We aim for a dead simple, modular design which is amenable to a clean and rigorous security analysis.

\paragraph{Overview of technical contributions.}
Under the peer-to-peer network assumptions of Section~\ref{sec:consistency}, the proposed construction in Section~\ref{sec:consensus} achieves a trustless, Nakamoto-based system which correctly reports on whether or not registered data are available during a given epoch.  Our construction maintains integrity without resorting to either distinguished nodes or a reputation scheme.  Any anonymous node can participate in the system with a deposit, and like Bitcoin, security improves as more honest and rational miners nodes join the network.  The main idea is straightforward: use Nakamoto consensus~\cite{Nak09b} to create a report of what's available and what isn't.

In the present system, every functioning Miner node stores and propagates each and every registered datum.  We do not attempt to ``shard'' data among various parties.   Rather than attempting an infinitely scalable solution, we instead content ourselves with designing a system where each node securely stores some orders of magnitude more data than ``on-chain'' storage permits.  Unlike a blockchain, which permanently stores all data forever, the present system releases data after its registration period ends, hence its storage space is reusable.  We shall further explore scaling methods in Section~\ref{sec:scalability}.

\medskip

Building on an independent blockchain offers some flexibility over Ethereum.  As noted earlier, Ethereum has limited storage space and its miners do not see external inputs.  Thus our data availability system requires some consensus beyond Ethereum mining.  Other approaches to trusted data feeds include data attestation via trusted hardware~\cite{towncrier} and, since the original publication of this manuscript, decentralized, on-chain voting via smart contracts~\cite{astraea,MVBV19}.

\section{Warm-up: fiat-crypto exchange rates}

Verifying an Ethereum transaction amounts to checking signatures, logical conditions, and self-evident mathematical truths.  Thus Ethereum miners, who are tasked with verifying these transactions, need only pay attention to local announcements on the blockchain.  One can, however, imagine that miners could agree on other kinds of objective, global facts, whose validity depends on time.  In this section, we outline a simple ``consensus computer''~\cite{LTKS15} application which extends beyond agreement on mathematical facts.

Consider the following concrete example.  Some decentralized applications require a bound on the exchange rate between a stable, fiat currency and a native, cryptocurrency token.  Since miners in traditional blockchains do not observe fiat exchanges, some other mechanism must necessarily feed this information to the blockchain.  Rather than relying on an authority, Teutsch and Reitwie{\ss}ner proposed to use a blockchain consensus protocol to agree on external exchange costs \cite[Section~5.5]{TR17}.  We shall now extrapolate on this idea and then, in Section~\ref{sec:consensus}, transmogrify it into a data availability protocol.

Network nodes agree on exchange rates using a variant of Nakamoto consensus~\cite{Nak09b}, the protocol underlying Bitcoin.  Rather than collating financial transactions into blocks, each Miner simply includes, in addition to a proof-of-work, the delimiters of a real interval allegedly containing the value of the current exchange rate.  A block, then, is valid if both the proof-of-work is correct and its interval contains the \emph{true} exchange rate accepted by a majority of other Miners.  By convention, Miners keep an eye on real exchange rates and mine on top of blocks they perceive as valid.

How large should the exchange rate interval be?  While in general, Miners might be able to agree, say, that the price of an ether lies between 650 and 700~USD, during moments of wild volatility or low liquidity, the market may not afford such precision.  For this reason, each Miner independently chooses the size of of the interval and receives a block reward inversely related to the size of the interval.  Smaller intervals have higher reward potential, however they also have a greater chance of being ignored by Miners who may perceive the associated block as invalid.

\section{The consistency problem} \label{sec:consistency}

Let us now return to our original problem.  We seek a decentralized system which correctly reports on data availability.  Our consensus protocol construction hinges on the following crucial assumption about the underlying peer-to-peer network.
\begin{consistency}
Either almost all nodes in the network can download a given datum, or almost none of them can.
\end{consistency}
The World Wide Web, for example, largely exhibits this property.  Either the website \url{truebit.io} is ``up'' and everyone in the world can see it, or else most people agree that the site is ``down'' regardless of their  Internet access point.  For the purpose of this exposition, we shall assume that a peer-to-peer network exists and exhibits this desired consistency property.   We shall not concern ourselves here with the construction of the peer-to-peer layer but rather explore a cryptoeconomic structure on top of it.

The Consistency Axiom above permits us to make the following well-defined notion.
\begin{defn}
Let $x$ be a datum.  If most nodes can see $x$, then $x$ is \emph{publicly available}.  If few nodes can see $x$, then $x$ is \emph{not publicly available}.
\end{defn}
Due to the gap afforded by the Consistency Axiom, the above definition covers all possible cases for all data.  We now introduce a secondary network assumption.
\begin{upload}
Any computer can join the network and, with high probability, propagate data into a publicly available state.
\end{upload}
Without the Upload Axiom, our Consistency Axiom might vacuously describe a network without any nodes.  A new upload need not become instantly available, however we shall assume a bounded lag on its propagation.  We conclude this discussion with one final requirement.
\begin{directory}
Nodes can efficiently determine whether or a datum~$x$ is publicly available from $\hash(x)$.  If $x$ is available, then $\hash(x)$ tells where to download $x$.
\end{directory}
Thus any node which wishes to upload or download a publicly available datum can use the datum's hash as a directory.  The IPFS~\cite{ipfs} filesystem, for example, satisfies the Directory Axiom.  Aside from exceptions described in Section~\ref{sec:analysis}, the system relies on Nakamoto consensus for security.

\section{A consensus protocol for data availability} \label{sec:consensus}

Under the axioms of Section~\ref{sec:consistency}, we now devise a decentralized system which indicates whether or not a registered datum is publicly available.  We assume familiarity with Nakamoto consensus~\cite{Nak09b}.  The network consists of two types of parties: Storers who wish to provably publish data and Miners who both confirm and guarantee availability of data.  Let $c$ be the \emph{network lag time}, or number of blocks to confirm a storage request for a datum and propagate it through the network.  For simplicity of presentation, we assume that this time bound suffices for all datums, regardless of size.  The Upload Axiom from Section~\ref{sec:consistency} now allows us to introduce the following upload interface.

\begin{description}

\item[Storer interface.]  
A Storer who wishes to publish a datum~$x$ broadcasts a \emph{registration} of $x$ to the network consisting of the following components:
\begin{enumerate}
\item {$\hash(x)$} which, by the Directory Axiom, doubles as an address for downloads,

\item the number of block epochs for which $x$ should be publicly available (excluding network lag time~$c$), and

\item a \emph{reporting fee} payable to Miners based on the size of $x$ and the registration duration specified in item~2.
\end{enumerate}

We say that datum~$x$ is \emph{registered at time~$t$} if the blockchain contains a registration for~$x$ that persists at block epoch~$t$.  Note that registered data may or may not be publicly available.  The Storer must propagate his registered datum, lest the network report it as unavailable.

\begin{defn} \label{def:report}
A \emph{report}, denoted $R(x)$, is a formal, plaintext assertion that datum~$x$ is publicly available.  A Miner who includes $R(x)$ in the $t^\text{th}$ block signals that, ``datum~$x$ is publicly available in both block epochs~$t$ and $t+1$.''  A set of reports~$S$ in block~$t$ is called \emph{complete} if for every registered datum~$x$, $R(x) \in S$ if and only if $x$ is publicly available in block epochs~$t$ and $t+1$.
\end{defn}
\end{description}

Using Definition~\ref{def:report}, we shall obligate each Miner who propagates a block claiming that~$x$ is publicly available at time~$t$ to further ensure that $x$ remains available at time~$t+1$.  We shall assume that a single miner propagating $x$ suffices to make the datum publicly available, hence an honest Miner can always ensure $R(x)$ holds and guarantee herself a block reward.  See ``Space economy'' in Section~\ref{sec:analysis} for further details.

\begin{description}
\item[Miner interface.]
Any Miner who wishes to join the network first identifies the ``longest'' blockchain, namely the one containing the greatest proof-of-work \cite{Nak09b}.  The miner obtains the longest blockchain from peer nodes and retraces it from its genesis block, observing each Storer transaction along the way in sequence, and noting which Storer requests are still registered at the present moment.  The Miner attempts to download, store, and propagate all currently registered data.  The Miner locally considers any data successfully downloaded to be publicly available.  She can determine validity of the current header block valid, according to the criteria below, after silently observing the chain for~$c$ epochs.  We assume that the initial, altrustic Miners in the system converge to a consistent world view during the first~$c$ blocks following the consensus genesis.

\begin{defn} \label{def:valid}
A \emph{valid} block at time~$t$ consists of the following elements:
\begin{enumerate}
\item the complete set of reports at time~$t$ (taking into consideration network lag time~$c$),
\item a collection of new, cryptographically signed Storer requests, and
\item a value \emph{nonce} witnessing a proof-of-work.  More specifically, the concatenation of the following components must hash to a small value:
\begin{enumerate}
\item the mining nonce \cite{Nak09b},
\item items 1.\ and 2.\ above.
\item a private key at which to receive the block reward and network fees, and
\item the hash of the previous block header.
\end{enumerate}
\end{enumerate}
\end{defn}

\begin{mainloop}
The protocol steps for the Miner now roughly follow Nakamoto consensus~\cite{Nak09b}.
\begin{enumerate}
\item In each block epoch, the first miner to find a valid block broadcasts it to the network and receives a block reward plus applicable reporting fees.

\item Upon verifying a new valid block, each Miner downloads, stores, and propagates all data registrations contained in the new block.  The Miner locally considers any data successfully downloaded to be publicly available and propagates the data as expediently as possible.

\item The mining race begins anew on top of the new block.
\end{enumerate}
\end{mainloop}
Miners always mine on the ``longest'' chain whose most recent $c$ blocks are all valid.  ``Longest'' here formally means the chain with the greatest proof-of-work, since block difficulty may change over time.  Miners need not store data which ceases to maintain registered status.
\end{description}
Unlike Bitcoin, the present blockchain construction has no notion of validity for individual ``transactions.''  Indeed reports are only valid as sets.  We remark that the validity of a block also depends on time because data presence can vary.  We thus realize a powerful application of the consensus computer~\cite{LTKS15} which both requires and permits agreement on facts external to mathematics.

\section{Security analysis} \label{sec:analysis}

In this section, we argue that the blockchain construction in Section~\ref{sec:consensus} accurately reports data availability.  More specifically, we argue that a report for a registered datum~$x$ appears in the $t^\text{th}$ block if and only if $x$ was publicly available during the $t^\text{th}$ block epoch.  In other words, the system records both when a registered datum is publicly available and when it isn't.  The ``Main Loop'' in Section~\ref{sec:consensus} dictates that Miners always broadcast valid blocks consisting of complete reports.  The security doubts we must resolve in order to establish our desired property above are twofold:
\begin{enumerate}
\item Can rational Miners and Storers gain by deviating from the consensus protocol?

\item Does the consensus withstand peer-to-peer layer failures?
\end{enumerate}
We shall assume that an adversary of the first type wishes to either:
\begin{enumerate}[label=1\alph*)]
\item convene a published report that some datum was publicly available when it actually wasn't, or
\item receive a block reward without actually committing resources to the network.
\end{enumerate}

Nakamoto consensus largely inhibits attacks of types 1a) and 1b).  The Consistency Axiom permits us to circumvent potential attack vectors of the second type, however precise protocol adaptations for handling these attacks depends on the specific implementation of the underlying peer-to-peer network.  It remains a crucial open problem to design and construct a peer-to-peer layer satisfying the three axioms in Section~\ref{sec:consistency}.

\paragraph{Data withholding.}
A Storer who registers a datum~$x$ could potentially take any of the following actions:
\begin{enumerate}
\item neglect to publish~$x$ itself,

\item delay his public reveal of~$x$ until the final moments of $c$-block propagation period, or

\item publish~$x$ at first but then hide this datum from the network.
\end{enumerate}
In case~1, Miners never see~$x$, unless it was otherwise publicly available, and hence they never report~$x$ in a block.  Since the blockchain witnesses no report of~$x$ as publicly available, this type 1a) attack fails.  Cases~2 and~3 do not impact functionality of the network, care of the Consistency Axiom.  Indeed, either $x$ becomes publicly available, in which case most Miners would report it as such, or else it would not be publicly available, in which case they would not.  Miners have incentive to propagate data (Section~\ref{sec:consensus}), which reduces the chances that $x$ could transition from publicly available to not publicly available during its registration period.  In short, the Storer cannot incite a fallacious or controversial report on the blockchain.

\paragraph{Space economy.}
Although the consensus protocol in Section~\ref{sec:consensus} obligates Miners to store and propagate data, a Miner might neglect this obligation in attempt to conserve resources in accordance with attack type~1b).  A Miner who fails to upload and propagate publicly available data risks orphaning his block as other Miners who cannot access~$x$ may perceive the block as invalid.  Similarly, the Miner who mines the subsequent block at time~$t+1$ bears responsibility for the availability of~$x$ until the next block epoch~$t+2$.  This overlapping storage requirement reinforces the Consistency Axiom by incentivizing Miners to download and propagate all registered, publicly available data.  Indeed, a Miner who witnesses but does not propagate data must rely on the faithfulness of other anonymous Miners and Storers to maintain validity of her reports in the successive mining race.

We remark that with some small probability, a Miner may not be able to see a publicly available datum and therefore might unknowingly propagate an invalid block.  This invalid block would then not be accepted by other Miners, who have the ``true'' world view.  As in Bitcoin, timely, well-intentioned blocks occasionally ``uncle,'' or never reach the blockchain.  Confirmed reports, however, remain consistent with publicly available data because the majority of Miners share correct perceptions.

\paragraph{Mining pools.} Miners often share computing resources and block rewards via \emph{pools} in order to collectively reduce income variance.  The pool \emph{operator}, who coordinates a pool's cooperation efforts, typically chooses the reports for all members in the pool.  Under such circumstances, pool members would rely on their pool operator to determine whether a datum is publicly available, thereby reducing the total number of eyes on the peer-to-peer network and weakening consistency.  A simple solution is to require all mining pool members to choose their own reports rather than relying on operators for selection.  To this end, the underlying consensus mechanism could, for example, mandate universal participation in SmartPool~\cite{LVTS16re}.

\section{Scalable implementation} \label{sec:scalability}

How much data could a system like the one proposed here actually monitor in practice?  While the actual capacity depends on the structure of the underlying peer-to-peer network, the present storage mechanism clearly has some finite limit for the same reasons that Bitcoin and Ethereum have bounded transaction volumes~\cite{LTKS15}.  The finite limit of this system should, however, greatly exceed what the system could securely store directly on the blockchain itself.  What about more storage --- does this construction scale?  Fortunately, two independent storage systems can store twice as much as a single one!

In theory, one can store an unlimited amount of data through system replications, however, if one wishes to use the availability of data in some root system, such as Ethereum, the root system's underlying consensus protocol must keep an eye on each individual, data availability system.  Suppose that a Task Giver in TrueBit \cite{TR17} were to provide a computational task whose off-chain input Solvers and Verifiers could not see (i.e.\ they witness only a hash on-chain).  Then the Task Giver could potentially provide a bogus solution to his own task, and neither Solvers nor Verifiers would have any recourse to object against the data unavailability unless the authoritative Judges, or miners, collectively expand their myopic, on-chain, world view.

Finally, we remark that maintaining the same level of proof-of-work security on two independent blockchains requires twice as much mining resources as a single blockchain.  Thus a hierarchical system such as Plasma~\cite{plasma}, in which the integrity of each child blockchain relies on proper function of its parent, could provide a useful scaling model for data availability.  Each leaf in Plasma's blockchain hierarchy would manage a modest amount of data, while the root system would only monitor data availability disputes for escalated situations.

\paragraph{Acknowledgements.} Thanks to Andreas Veneris for useful comments which are reflected in the updated version.

\bibliographystyle{plain}
\bibliography{decentralized_oracles}

\end{document}